\documentclass[preprint]{aastex}
\usepackage{CJK}
\usepackage{amssymb}
\usepackage{amsmath}
\usepackage{graphicx}
\usepackage{graphics}
\usepackage{amsfonts, color, hyperref, epsfig}

\newcommand{\lya}{Ly$\alpha$}

\newcommand{\hl}{H\,{\footnotesize I}}
\def\J1259{\mbox{SDSS~J1259+6212}}

\begin{document}
\title{A Strange  EUV Emission: Scattered Continuum in the Lymann Limit Absorption Edge toward the Quasar SDSS J125903.26+621211.5?}
\author{Xiang Pan\altaffilmark{1,2}, Shaohua Zhang\altaffilmark{2}, Hongyan Zhou\altaffilmark{1,2}, Xiaoyi Xie\altaffilmark{2}, Xiheng Shi\altaffilmark{2}, Peng Jiang\altaffilmark{2}, Ning Jiang\altaffilmark{1} and Weimin Yuan\altaffilmark{3}}
\affil{$^1$CAS Key Laboratory for Research in Galaxies and Cosmology, Department of Astronomy, University of Sciences and Technology of China, Hefei, Anhui 230026, China\\
$^2$SOA Key Laboratory for Polar Science, Polar Research Institute of China, 451 Jinqiao Road, Shanghai 200136, China; zhangshaohua@pric.org.cn, zhouhongyan@pric.org.cn\\
$^3$CAS Key Laboratory of Space Astronomy and Technology, National Astronomical Observatories, Chinese Academy of Sciences, Datun Rd 20A, Beijing 100012, China}
\begin{abstract}
We identified a  peculiar proximate sub-damped \lya\ absorption system (sub-DLA) at z=3.234 with a neutral hydrogen column density of $N_{\rm \hl} =10^{19.70\pm0.15}~\rm cm^{-2}$ toward  the quasar SDSS J125903.26+621211.5 in two epoch optical spectra of the Sloan Digital Sky Survey. We detected \lya\ residue in the proximate sub-DLA trough at $> 8\sigma$ level. To our surprise,  extreme ultraviolet (EUV) continuum emission is significantly ($> 4\sigma$) detected in the corresponding Lymann limit absorption edge at both of the FUV and NUV bands by the Galaxy Evolution Explorer.
The high neutral hydrogen column density should allow a negligible transmission of both the \lya\ line photons and EUV continuum photons due to the high optical depth of the gas.
The possible scenarios of foreground galaxy contamination, partial coverage, emission from the quasar host galaxy, and extended \lya\  emission are excluded in turn,
and we speculate that the residual \lya\ and EUV emission is due to photons scattering (broad \lya\ and the continuum emission) of  electrons residing at a spatial scale larger than that of the proximate sub-DLA.
Electron scattering is wavelength-independent, the scattered light is therefore a copy of the incident spectrum that might originate in the accretion disk.
With the assistances from the neutral hydrogen absorbers as the ``natural coronagraph'' and the scatterers as the ``natural mirror'',
we yielded a very hard EUV spectral index of $\alpha = 0.65\pm 0.25$ ($F_{\nu}\propto \nu^{\alpha}$), consistent with the standard picture of the locally heated accretion disk in the inner EUV-emitting radii, as well as in the outer near-infrared-emitting radii suggested by Kishimoto et al. (2008).
\end{abstract}

\maketitle

\section{INTRODUCTION}
Hydrogen, as the most abundant element in the universe, widely appears in the galaxy and the intergalactic medium. Neutral hydrogen clouds with  high column densities are usually considered to be the high-redshift progenitors of present-day galaxies (e.g., Wolfe et al. 1986; Nagamine et al. 2004; Prochaska et al. 2005). Thus, the study of neutral hydrogen plays essential roles in understanding the formation and evolution of galaxies. Because active galactic nuclei (AGNs), particularly quasars, are generally luminous, high column density neutral hydrogen within whether the cosmological intervening or quasar-associated environments, would be traced in absorption under the illumination of background radiation from quasars. As the column density increases, the absorption features are  the \lya\ forest absorber ($N_{\rm \hl}<10^{17}~\rm cm^{-2}$), the Lyman limit system (LLS; $10^{17}<N_{\rm \hl}<2\times10^{20}~\rm cm^{-2}$), and the damped \lya\ absorption system (DLA; $N_{\rm \hl}>2\times10^{20}~\rm cm^{-2}$), respectively (reviewed in Wolfe et al. 2005).

In the intervening system, the emission of the background quasar is completely absorbed in the DLA and sub-DLA (also called as Super-LLS; $10^{19}<N_{\rm H I}<2\times10^{20}~\rm cm^{-2}$) trough, as well as the flux at rest wavelengths of $\lambda < 912~\rm \AA$.
Weak residual \lya\ emission in individual sources (e.g., M{\o}ller et al. 2004; Fynbo et al. 2010; Kulkarni et al. 2012; Jiang et al. 2016) and  the large-sample composite spectrum (Noterdaeme et al. 2014) are detected. Faint starlight emerges with the assistance from the neutral hydrogen absorbers as the natural ``coronagraph'', thus it is considered to be the \lya\ emission from vigorous star formation in the host galaxy of the absorbers (e.g., Prochaska et al. 2005, 2014). However, Cai et al. (2014) attributed residual flux in the composite DLA troughs to FUV stellar light from the quasar host galaxies, since the residual flux in the dark troughs of the composite DLA spectra is strongly correlated with the continuum luminosity of the background quasar, while uncorrelated with DLA column density or metallicity.
In the quasar-associated system,   a slightly larger amount of residual fluxes likewise are found in prominent DLA troughs (Pan et al. 2017), while the reasons  are much more complicated than those in the intervening system. The residual \lya\ emission would be interpreted as partly obstructed by the absorber, as associated with star formation activity, as scattered \lya\ photons from the quasar, or even as luminous, spatially extended \lya\ blobs (e.g., Weidinger et al. 2004; Hennawi et al. 2009;  Finley et al. 2013; Fathivavsari et al. 2015, 2016, 2017; Jiang et al. 2016; Pan et al. 2017).

In the present paper, we report the detection of significant residual \lya\ emission in the peculiar prominent sub-DLA trough toward SDSS J125903.26+621211.5 (hereafter \J1259). In addition, intriguingly, the Galaxy Evolution Explorer (GALEX; Morrissey et al. 2007) present the extreme ultraviolet (EUV) emission passing the FUV and NUV filters beyond the Lyman limit at $>4\sigma$ levels, which should be completely absorbed by the optically thick neutral hydrogen cloud.
 These observations raise questions about the origin of this unusual residual EUV and \lya\ emission. 
This paper is organized as follows: Section 2  presents the detailed analysis of the absorption system, and the origin of the residual EUV and \lya\ emission as well as applications are discussed in Section 3. The adopt cosmological quantities  are H$_{\rm 0}$ = 70 km s$^{-1}$Mpc$^{-1}$, $\Omega_{\rm m} = 0.3$, and $\Omega_{\Lambda} = 0.7$.

\section{OBSERVATIONS AND DATA ANALYSIS}
\J1259\ is a bright ($m_{r_{\rm SDSS}}=18.19\pm0.01$) high-redshift quasar at $z_{\rm em}=3.2340\pm0.0002$ identified by the Sloan Digital Sky Survey (SDSS; York et al. 2000). Two optical spectra were acquired with the SDSS spectrograph at Feb. 25, 2008 and with the SDSS-III/Baryon Oscillation Spectroscopic Survey (BOSS) spectrograph (Dawson et al. 2013) at Apr. 16, 2014, respectively.
When we carefully compared the SDSS spectrum with the BOSS spectrum in continuum and emission/absorption lines, no significant variation in either of these spectral features are detected between the time interval of $\sim6$ years in the observer's frame.  Figure 1 (a) presents the BOSS spectrum and flux noise in black and pink.

The BOSS spectrum of \J1259\ ranges from 843 to 2438 \AA\ in the quasar's rest-frame. It enables us to search for the associated metal absorber system corresponding to the neutral hydrogen Lyman-series absorptions.
In Figure 1 (a) and (c), the spectrum and the normalizations show an quasar-associated absorption system with abundant metal lines of \ion{O}{6}\ $\lambda\lambda1031,1037$,
\ion{N}{5}\ $\lambda\lambda1238,1242$, \ion{C}{4}\ $\lambda\lambda1548,1550$, \ion{Si}{4}\ $\lambda\lambda1391,1402$, \ion{C}{3}\ $\lambda977$, \ion{N}{3} $\lambda989$, \ion{Si}{3} $\lambda\lambda1193,1206$, and etc. (marked by gray characters).
We also noticed the overdensity of absorptions blueward of the strong \lya\ trough. Some of them have been confirmed as high-order Lyman lines or metal absorptions from the same system, such as \ion{O}{6} and \ion{Si}{3}. But others remain unidentified, though we prefer to attribute them to \lya\ from intervening absorbers, since they are weak and narrow, with widths comparable to the instrumental profile. As shown in Figure 2, the metal absorption lines present asymmetrically irregular profiles in velocity space of the quasar's rest frame, which seems to have some velocity structure. But the intermediary resolution of the SDSS and BOSS spectrograph is not enough to resolve the structure, then we calculated the absorption-weighted average velocity of the absorption lines (generally used for broad absorption lines; Trump et al. 2006; Zhang et al. 2010, 2014; Liu et al. 2015) as the absorption redshift (dotted lines in Figure 2). Relative to the quasar's rest frame, the metal absorption lines are blueshifted with a velocity of $\sim170~\rm km~s^{-1}$.

In the blue end of the spectrum, the fluxes at rest wavelengths of  $\lambda \lesssim 912~\rm \AA$ are deeply suppressed, suggesting an optically thick neutral hydrogen at the Lyman limit. Furthermore, the damped \lya\ absorption trough reveals that the associated absorber has a high column density of neutral hydrogen (Figure 1 (b)). The higher-order Lyman-series absorption from Ly$\beta$ to even Ly$\iota$ are credibly identified despite contamination from the \lya\ forest (Figure 1 (c)). In Garnett et al.(2017), a intervening DLA absorber toward \J1259\ is reported at $z_{\rm abs}=2.6108$, but actually the trough they detected is the absorption of \ion{O}{6}\ $\lambda1037$ (marked in Figure 1 (c)).
Moreover, it is easily found that the bottom fluxes in the higher-order Lyman-series
troughs are almost to vanishing point (much less than the noises), as well as the saturated \ion{C}{3} absorption line and \ion{C}{4} doublet, suggesting that the quasar's nuclear region of \J1259\ is fully covered by the associated absorber.

In order to normalize the observed spectra, we should create a absorption-free spectrum for \J1259.
A single power-law slope is firstly estimated from continuum windows listed in Forster et al. (2001), which are not seriously contaminated by emission lines to represent the nuclear continuum.
The profiles of \lya, \ion{N}{5} and other emission lines are reconstructed by multi-Gaussian components. The single Gaussuian component has no physical meaning, the modeling profile is simply to recover the unabsorbed spectrum.
The natural extension of the power-law continuum is applied for the pixels at the shortward of the Lyman edge. This phenomenological spectral-decomposition method are generally used in the measurements of the broad emission line (e.g., Dong et al. 2008; Shen et al. 2011) and the unabsorbed spectrum recuperation (e.g., Zhang et al. 2010, 2015).
It is worth noting that the emission-line profile of Ly$\alpha$ is primarily constrained from the red wing of Ly$\alpha$ and \ion{N}{5} lines because of the serious absorption of Ly$\alpha$ and metal lines at the blue wing of Ly$\alpha$.
The final absorption-free spectrum is overplotted by red curves in Figure 1 (a) and (b).
We measure the column density of neutral hydrogen using Voigt profile fitting to the \lya\ trough, and then check the prediction of the reasonable for the higher-order absorptions.
Before modeling is applied, pixels affected by incident absorption features are carefully masked, and the redshift of the DLA trough is fixed to that of the metal absorption system.
Profile fitting is  carried out with an IDL MPFIT procedure and the results in a column density of $N_{\rm \hl} =10^{19.70\pm0.15}~\rm cm^{-2}$ are shown by blue curves in Figure 1 (b). The best-fit model well reproduces the damped wings of \lya\ and  other higher-order Lyman-series, while there are significant residual fluxes at nine pixels in the center of the \lya\ trough (yellow area).
 With such a high neutral hydrogen column density, the higher-order Lyman-series absorption lines should be saturated (zero fluxes or fluxes submerged in the errors at the positions of the line centers). That is consistent with the observed spectra except the troughs of Ly$\theta$ and Ly$\iota$ affected by large spectral fluctuations (Figure 1(c)). The average  flux density (of the residual emission) is $\rm (12.53\pm1.51)\times10^{-17}~erg~s^{-1}~cm^{-2}~\AA^{-1}$.

\J1259\ was also reported as a far-UV bright quasar (Syphers et al. 2009). The AB magnitudes released on the GALEX website through GALEXVIEW are $22.08$ mag for FUV band (centered at $\rm 364~\AA$ in the quasar's rest frame) and $22.38$ mag for NUV band (centered at $\rm 537~\AA$), respectively. Since \J1259\ is very faint in UV, the image is easily polluted by the UV emission from other surrounding weak sources and the measurement is probably influenced by the selection of aperture size.
However, from the SDSS images, one can find that the nearest suspected dim object is $\sim 10.4$ arcsec from SDSS J1259+6212.
A specialized photometry with an aperture radii of 6 arcsec is thus carefully performed by extracting the FUV and NUV images from the GALEX All-Sky Imaging Survey (AIS).   Fortunately, The resultant  FUV and NUV  AB magnitudes are $21.91\pm0.26$ and $22.26\pm0.21$ mag, respectively.
The values are little different from those released by the pipeline, and highly consistent with the flux measurements presented in Syphers et al. (2009), which are 7.35 and $\rm2.62 \times10^{-17}$\mbox{erg s$^{-1}$ cm$^{-2}$ $\rm\AA^{-1}$} in the observational frame (also see Figure 1(a)).
For the Lyman absorber with $N_{\rm \hl} \approx 10^{19.70\pm0.15}~\rm cm^{-2}$, the optical depth at $\rm 320 - 420~\AA$ (FUV band in the quasar's rest frame) would exceed 15, and at $\rm 420 - 660~\AA$ (NUV band in the quasar's rest frame) the optical depth would exceed 35, which should allow few background EUV photons remaining unabsorbed.

Meanwhile, as discussed above, there might be intergalactic medium (IGM) transmission in the the blueward of the sub-DLA \lya\ trough.
These IGMs would contribute to the total optical depth at the rest NUV and FUV bands, but a quantitative analysis would not be easy due to the limited SNR and resolution of the SDSS spectra. Anyhow, the optical depth we derived according to the H$^0$ column density of the sub-DLA system present the lower limit for the total optical depth. And the introduction of the Lyman continuum absorption from IGM would not affect our conclusion.

\section{DISCUSSION}
Utilizing the optical spectra from the SDSS and SDSS-III/BOSS, we analyze in detail the residual \lya\ emission in the center of hydrogen absorption trough toward \J1259. The abundant metal transitions indicate the absorption system is quasar-associated with a blueshifted velocity of $\rm \sim 170~km~s^{-1}$ in the quasar's rest frame.
A neutral hydrogen cloud with the column density of $N_{\rm \hl}= 10^{19.7\pm0.15}~\rm cm^{-2}$ in proximity to the quasar is identified via the detections of an associated sub-damped \lya\ trough, high-order Lyman series absorptions, and the Lyman limit absorption edge in the optical spectra. The residual \lya\ emission has a average flux density of $\rm (12.53\pm1.51)\times10^{-17}~erg~s^{-1}~cm^{-2}~\AA^{-1}$, and the total luminosity of \lya\ residues in the damped trough reaches to $L_{\rm Ly\alpha} =(1.65\pm0.24)\times10^{44}\rm ~erg~s^{-1}$.
Meanwhile, intriguingly, the GALEX AIS presents the significant EUV continuum residues beyond the Lyman limit at $> 4\sigma$ levels, the AB magnitudes are  $21.91\pm0.26$ and $22.26\pm0.21$ mag, respectively.
Despite the faint UV magnitudes, the UV brightness is nonetheless dramatically high, considering the exist of the proximate sub-DLA system. The existence of the residual EUV and \lya\ emission raises interesting questions as what the significant residues are and what the physical mechanism behind this phenomenon is.

Generally speaking, there are only a few scenarios for the significant residues in a sub-DLA system, i.e., foreground galaxy contamination, partial coverage, emission from the quasar host galaxy, extended Ly$\alpha$ emission, and scattered light from the accretion disk.
The first scenario to consider is the impact of foreground galaxy.
Foreground galaxy contamination is often a challenge in searches for Lyman continuum emitter galaxies (e.g., Nestor et al. 2011, 2013; Siana et al. 2013, 2015; Vanzella et al. 2010, 2012).
The advantage of the quasar spectrum is that we would detect the indications of the metal absorption lines (e.g, \ion{Mg}{2}, \ion{Zn}{2} and \ion{Fe}{2}) superimposed on the continuum of the quasar (for the interlopers with $z > 0.4$), and the emission fluxes in absorption troughs of high-order Lyman series (for the blue galaxies with strong starformation and appropriate redshifts).
Furthermore, the SDSS five-band photometry would detect the interloper's images (for the interlopers with low redshifts). However, it is clear that the spectroscopy and imaging observations of SDSS J1259+6212 do not supported the foreground galaxy scenario.

The second scenario to consider is that the nucleus region of \J1259\ is imperfectly blocked by the absorber. However the almost completely absorbed bottoms of saturated troughs of high-order Lyman series (Figure 1 (c)), \ion{C}{3} and \ion{C}{4} (Figure 2) easily rule out this possibility. In particular, the absorption lines of \ion{C}{4} $\lambda\lambda1548,1550$  coincides with the strong broad emission line, their zero fluxes in the bottom indicate that the broad line region is completely obscured.
If the \lya\ residues may be due to imperfectly blocked broad line region emission, we would at least observe the residual emission in the \ion{C}{4} trough.
From this point of view, \J1259\ is very different from other proximate cases with residual \lya\ emission, e.g., SDSS J082303.22+052907.6, SDSS J011226.76-004855.8, and SDSS J113341.29-005740.0. In these three cases, the \lya\ residuals are considered as the emission from the narrow/broad emission line regions because of the partly
eclipsed effect of the coronographs  (Fathivavsari et al. 2015, 2016, 2017).

The third scenario to consider is emission from the quasar host galaxy.
The approach of using the residual flux in DLA troughs to observe the \lya\ emission associated with star formation in the host galaxies has been studied extensively (e.g., Kulkarni et al. 2006; Fynbo et al. 2010; Finley et al. 2013; Noterdaeme et al. 2014). Using the Kennicutt (2008) calibration $SFR {\rm (M_{\sun}~yr^{-1})} = L_{\rm H\alpha}/ 1.26\times10^{41}\rm~ erg~s^{-1}$ and assuming an intensity ratio of $\rm Ly\alpha/H\alpha=8.3$ for the case B recombination, we estimate a star formation ratio of $\rm 150\pm23~M_{\sun}~yr^{-1}$ based on the luminosity of the residual \lya\ emission.
That is much higher than  estimates of  the SFR in quasar host galaxies ($\rm \sim 9~M_{\sun}~yr^{-1}$; e.g., Ho 2005; Cai et al. 2014) and in the high redshift Lymann break galaxies and \lya\ emitters ($\rm \sim 30~M_{\sun}~yr^{-1}$; e.g., Shapley et al. 2003; Erb et al. 2006; Gronwall et al. 2007). To further check whether the residual \lya\ emission originated from the  star formation, we compare the colors  (the flux density ratios: FUV/NUV and NUV/\lya) of the residues with those of the galaxies with active star formation.
The comparison galaxies are selected from the starburst original 1999 dataset (Leitherer et al. 1999). The spectral energy distributions (SEDs) with ``\emph{Topic} - stellar continua'' and ``\emph{Quantity} - stellar emission only'' ($\rm Fig.~ 7-12$ in the Starburst99 website http://www.stsci.edu/science/starburst99/docs/table-index.html) are used to compute galaxy's colors. The starlight models present in a homogeneous way for five metallicities between $Z= 0.040$ and 0.001 and three choices of the initial mass function. The age coverage is from 1 Myr to 1 Gyr, and both star formation law (instantaeous and continuous) are contained. The SEDs placed at the quasar's redshift are are extracted the flux densities at the effective wavelengths of the FUV/NUV filters and 1215.6 \AA. Figure 3 (a) shows that \J1259\ seriously deviates from the star forming galaxy group.
At the wavelengths of a few hundreds of Angstrom, \J1259\ has a much bluer slopes than the bluest galaxy continuum.
In Fathivavsari et al. (2018), strong and narrow \lya\ emission is reported in 155 eclipsing damped \lya\ systems, which is certainty revealed as the narrow emission line from the host galaxy. However, the residual \lya\ emission of \J1259\ fills in the sub-DLA trough rather forming a marrow peaked profile, and its luminosity is also higher than those from the Fathivavsari et al. eclipsing DLA sample. Thus the host galaxy scenario is questioned on the other hand.

The fourth  scenario to consider is extended \lya\ emission.
From the integral field unit (IFU) observations, the number of extreme \lya\ nebulae detected around bright quasars is growing (Cai et al. 2018 and references therein). Borisova et al. (2016) found a hundred percent detection rate of \lya\ nebulae around 17 $3<z<4$ quasars, and extended \lya\ halos are also extremely common around \lya\ emitter galaxies (Wisotzki et al. 2016; Leclercq et al. 2017). \lya\ emission of which can be produced from collisional excitation of hydrogen in the `cold flow' model of galaxy formation (Haiman et al. 2000; Fardal et al. 2001), and from ambient warm ionized gas illuminated by obscured AGNs and starbursts (e.g., Chapman et al. 2004; Dey et al. 2005; Scarlata et al. 2009), as well as outflowing superwinds (e.g. Ohyama et al. 2003; Bower et al. 2004). In this case, large, spatially extended region of luminous line emission will be presented, reaching sizes of order 100 kpc and line luminosities  of $\rm \sim 10^{44}~erg~s^{-1}$.
Interestingly, the residual \lya\ emission of SDSS J1259+6212 has a comparable luminosity to the sample from Borisova et al. (2016). But it is collected by the fiber with only 2 arcsec diameter, which corresponds to the physical scale of $\rm \sim 15~kpc$, approximate ten times smaller than the extended \lya\ emission regions. Meanwhile, the residual \lya\ emission is not significantly different from that in the SDSS spectrum obtained through 3 arcsec fiber (Figure 1 panel (b)).
Based on the surface brightness distribution of the \lya\ nebula (Figure 5 of Borisova et al. 2016), the residual \lya\ in the SDSS spectrum would be $\sim 30\%$ larger than that in the BOSS spectrum. Furthermore, the \lya\ nebulae (arising from recombination radiation or collisional excitation) cannot produce the EUV continuum emission.  This evidence implies that the residual EUV and \lya\ emission in SDSS J1259+6212 do not meet the scenario of extended \lya\ emission.

Finally, as the above discussion of the origin of the residual \lya\ emission suggests, scattered \lya\ photons from the quasar become the only option left.  Furthermore, photon scattering can also account for the residual EUV emission. 
Here it is worth looking back at the energy source of AGNs. AGNs are the most luminous objects in the universe. The energy results from an putative optically thick accretion disk located in the center, which is heated locally by the dissipation of gravitational potential energy with the surrounding gas falling (e.g., Shields 1978; Malkan \& Sargent 1982). This leads to the well-known observable spectral feature of the AGN's continuum in the UV/optical decade ($\sim 0.01-0.4$ $\mu$m), i.e., the ``big blue bump''. In the case of local blackbody emission assumption, the effective disk temperature $T$ is a function of radius $r$ as $T\varpropto r^{-3/4}$, the energy per unit frequency of the disk spectrum would be approximately increasing with frequency like $\nu^{+1/3}$ in a wide wavelength range from the UV/optical to the near-infrared (NIR; $\sim 1-2$ $\mu$m) (e.g., Shakura \& Sunyaev 1973).
 Power-law fits to the observed UV/optical continuum typically yield much redder slopes than the prediction (e.g., Neugebauer et al. 1987; Cristiani \& Vio 1990; Francis et al. 1991; Zheng et al. 1997; Vanden Berk et al. 2001£» Lusso et al. 2005),
and in particular, Lusso et al. suggests that the continuum slope shows a break at $\rm \sim 912~\AA$. 
Considering the complexity of absorption (especially extinction), it is not entirely surprising to us that the power law index at $\rm \lambda < 912 ~\AA$ is  softer   than that at longer wavelength, although the slope has been operated with a ``state-of-the-art'' correction. The intrinsic EUV emission of AGN is an interesting topic but difficult to observe. Identifying particular ``lucky'' individual cases that are extremely blue in the EUV wavelengths, such as J1259+6212, and undertaking a statistical analysis (beyond the scope of this paper) deserves further exploration as a new way to investigate the disk continuum. In one approach to characterize the accretion disk,   Kishimoto et al. (2008) provided a weight of evidence that, at least for the outer NIR-emitting radii, the locally heated disk is approximately correct based on the observations of polarized light in the infrared. In their work, the weighted mean of the measured slopes from the polarized NIR spectra is $\alpha=+0.44\pm0.11$, consistent with the predicted dependence. It reminds us that  observation of scattered AGN radiation may provide an opportunity to reliably examine the emission of the accretion disk.

Under this assumption, the residual EUV emission of \J1259\ is consisted of the scattered photons of the continuum, and then the residual EUV is also, possibly, an opportunity for us to directly examine the inner UV-emitting radii of the accretion disk.
The fitting for the FUV and NUV fluxes by a shape of power-law form presents a rapid decrease in $\nu F_{\nu}$ with the spectral index of $\alpha=+0.65\pm0.25$ (displayed by dashed line in Figure 3 (b)), which is approximately consistent with  the $F_{\nu}\varpropto\nu^{+1/3}$ shape. That seems to confirm the assumption that the residual flux comes from the scattered  photons from the quasar.The residual \lya\ flux is ten times higher than the extended slope (dashed line), but the value actually contains the contributions of the continuum and \lya\ broad line.
In the reconstruction of the absorption-free spectrum (red curve in Figure 1), we obtain the underlying-continuum and the emission line fitting results (including \lya\ line), respectively. Then we can easily estimate the flux ratio of the continuum and \lya\ line at the velocity of $\rm -170 ~km~s^{-1}$. When we assume that the fraction of the continuum in the residual flux in the sub-DLA trough is equal to that of the absorption-free spectrum, the scattered continuum is estimated and overplotted by open square in Figure 3 (b), and it is near  to the $F_{\nu}\varpropto\nu^{+1/3}$ line.

We do not know whether the scatterers are electrons or dust grains, and this could be probed with the possible polarimetric observations of SDSS J1259+6212. The two spectroscopic diagnostics are the scattering efficiency and the polarization
fraction of the scattered light (e.g., Kishimoto et al. 2001 and references therein). Electron scattering produces wavelength-independent scattering efficiency and polarization fraction, the scattered light therefore copies the spectrum originating in the region interior to the scattering region, however, dust scattering generally is wavelength-dependent,
and the strongest wavelength dependence of scattering efficiency and polarization fraction is in the EUV wavelengths.
Following the suggestion of Kishimoto et al. (2008), if electrons act as the scatterers in SDSS J1259+6212, then the measured slope suggests that the standard picture of the disk in the inner EUV-emitting radii being optically thick and locally heated is approximately correct.
The study of SDSS J1259+6212 demonstrates that, we may finally snoop the accretion disk with the assistance from the proximate lyman absorption troughs as natural ``coronagraph'' and the scatterers as the natural ``mirror''.
By inspecting the $\sim$ 600 eclipsing DLAs in Fathivavsari et al. (2018), we do find EUV residual fluxes in $\sim$ 10 objects. However, the residual fluxes are too weak for either an assured detection or measurements of EUV spectral index, partly because nearly all of the sample quasars ($> 99\%$) are not as bright as SDSS J1259+6212. The EUV residual fluxes in a sample of bright quasars with proximate DLA/sub-DLA might tell us further stories about the innermost region of AGN accretion disks, which will be presented in our future works.

\acknowledgments{The authors appreciate the enlightening suggestions from the anonymous referee, which helped to improve significantly the quality of this paper. This work is supported by National Natural Science Foundation of China (NSFC-11573024, 11473025, 11421303) and  National Basic Research Program of China (the 973 Program 2013CB834905). Funding for SDSS-III has been provided by the Alfred P. Sloan Foundation, the Participating Institutions, the National Science Foundation, and the U.S. Department of Energy Office of Science. The SDSS-III Web site is http:// www.sdss3.org/.}

\figurenum{1}
\begin{figure*}[tbp]
\epsscale{1.0} \plotone{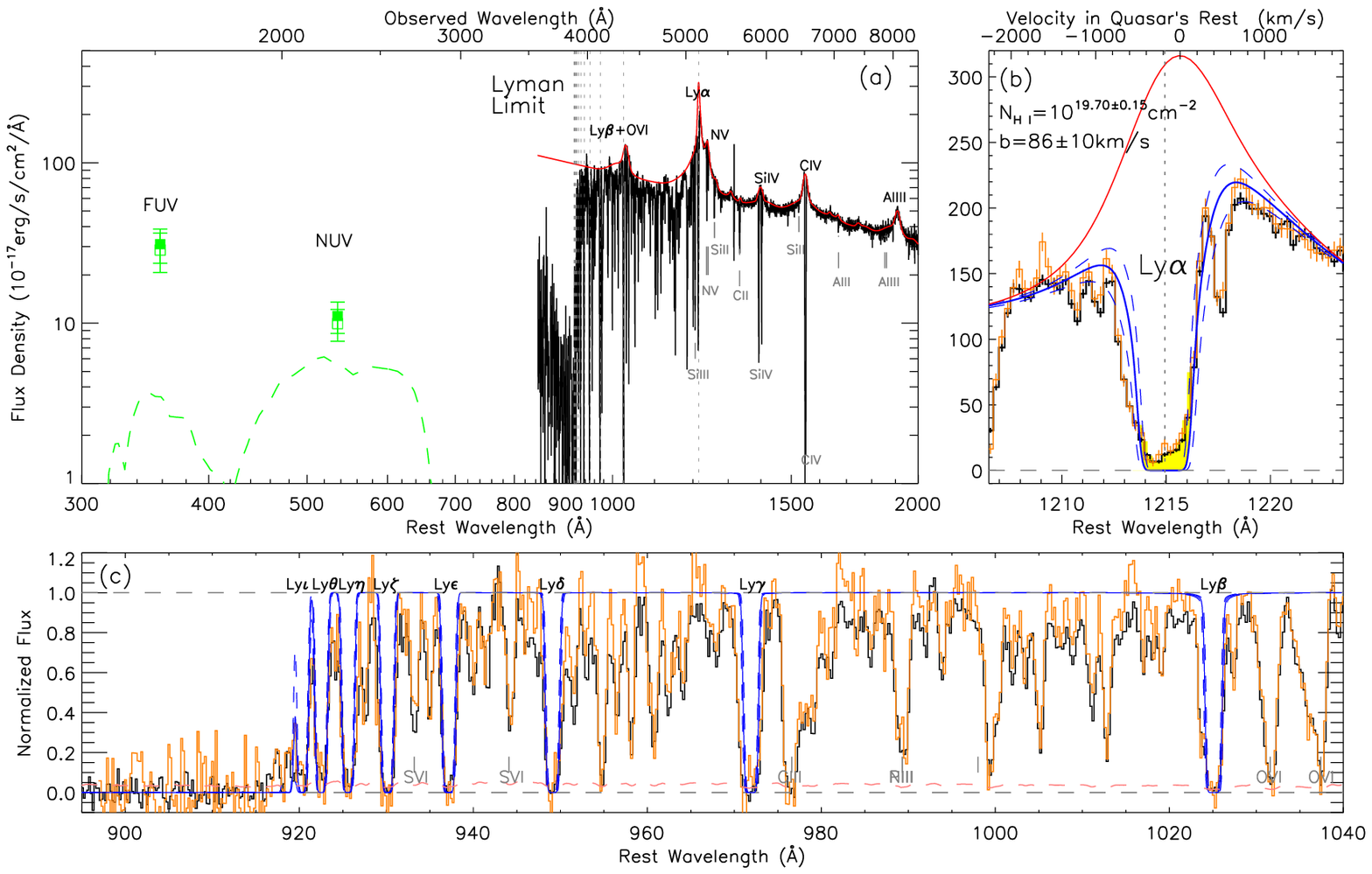}
\caption{Panel (a): BOSS spectrum and GALEX fluxes of \J1259. If not stated specifically, the rest wavelength in all figures of this paper corresponds to the rest-frame of the quasar at $z_{\rm em}=3.2340$. The emission lines and the strong absorption lines are labeled with dark and gray characters, respectively. The red solid curve means the normalization of the spectrum (unabsorbed spectrum). The transmission curves ($\times 10$) of the GALEX FUV and NUV filters are shown by green dashed curves.
The green squares are the FUV and NUV flux densities from the specialized aperture photometry (solid) and GALEXVIEW (open).
Panel (b): Analysis of Ly$\alpha$ absorption in the spectrum of \J1259. Voigt profile fitting to the damped \lya\ trough in blue: $N_{\rm HI}=10^{19.70\pm0.15}~\rm cm^{-2}$, the residual flux in the line center is labeled in yellow. The SDSS spectrum is also presented in orange.
Panel (c): Modeling of the Lyman-series absorption and the Lyman limit overplotted in normalized spectra (BOSS in black and SDSS in orange). The dashed pink curve is the normalized flux noise in the spectrum.
}\label{f1}
\end{figure*}

\figurenum{2}
\begin{figure*}[tbp]
\epsscale{1.0} \plotone{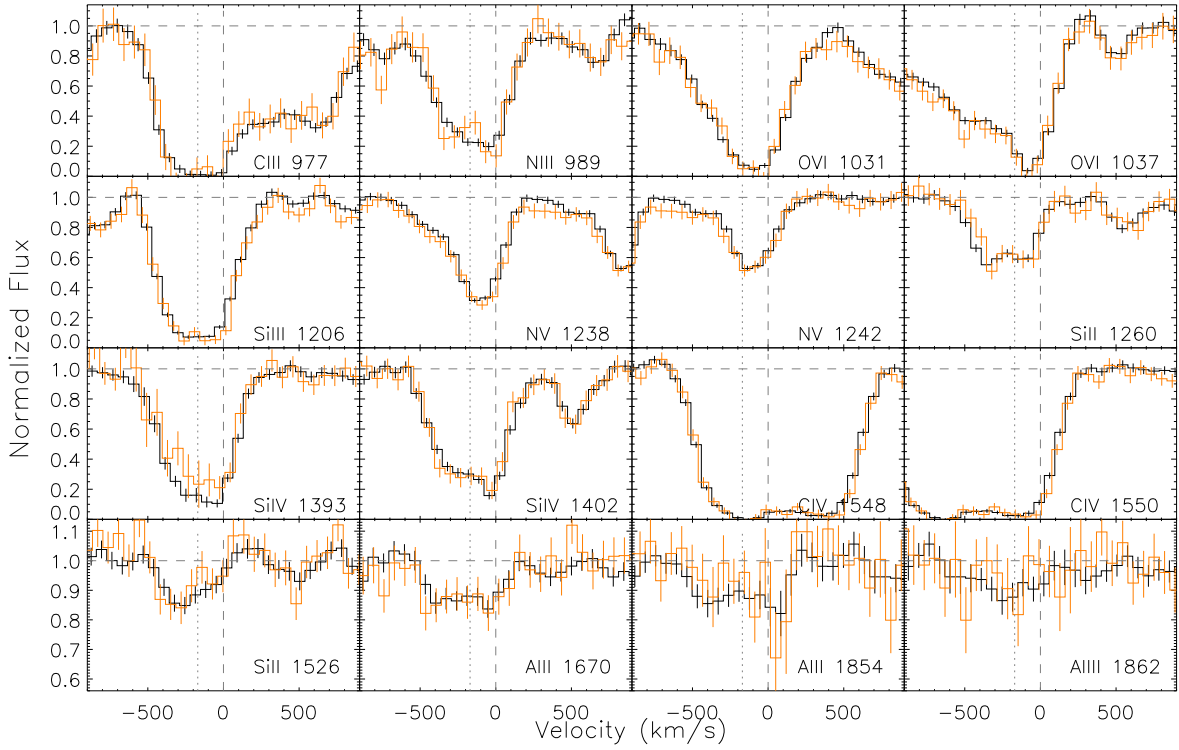}
\caption{Normalized metal absorption lines for \J1259 BOSS (black curves) and SDSS (orange curves) spectra in velocity space of the quasar's rest frame. The grey dotted lines at $v \sim -170~\rm km~s^{-1}$ indicate the absorption-weighted average velocity of the absorption lines.}\label{f2}
\end{figure*}

\figurenum{3}
\begin{figure*}[tbp]
\epsscale{1.0} \plotone{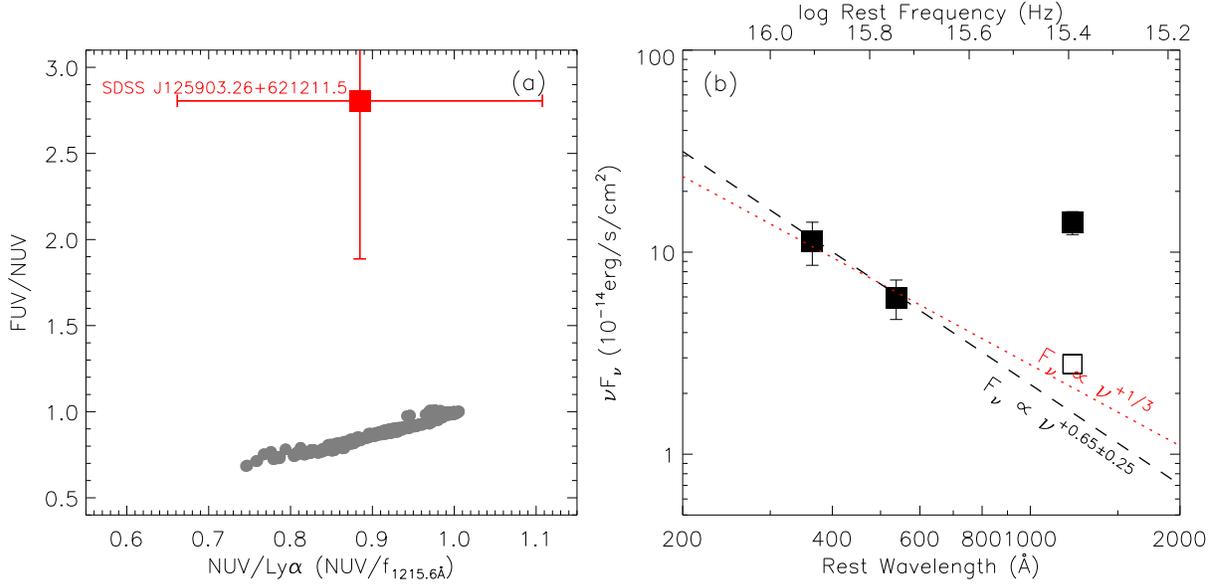}
\caption{Panel (a): Color (flux density ratio) comparison of the residual EUV and \lya\ emission of SDSS J1259+6212 (red square) and stellar continua
 in Starburst99 (gray circles). Starlight models are firstly placed at the quasar's redshift ($z_{\rm em}=3.2340$) and then are extracted the flux densities at the effective wavelengths of the FUV/NUV filters and 1215.6 \AA. \J1259\ seriously deviates from the star forming galaxy group. Panel (b): Power-law form fitting of the residual EUV emission (dashed line). The residual \lya\ emission is ten times higher than the extended slope, the flux actually contains the contributions of the continuum and \lya\ broad line. The devised continuum based on the line/continuum ratio of our absorption-free spectral template is overplotted by open square. The red dotted line is the linear fitting result with a fixed spectral index of $\alpha=+1/3$.}\label{f3}
\end{figure*}


\begin{thebibliography}{}
\bibitem[Borisova et al.(2016)]{2016ApJ...831...39B} Borisova, E., Cantalupo, S., Lilly, S.~J., et al.\ 2016, \apj, 831, 39
\bibitem[Bower et al.(2004)]{2004MNRAS.351...63B} Bower, R.~G., Morris, S.~L., Bacon, R., et al.\ 2004, \mnras, 351, 63
\bibitem[Cai et al.(2014)]{2014ApJ...793..139C} Cai, Z., Fan, X., Noterdaeme, P., et al.\ 2014, \apj, 793, 139
\bibitem[Cai et al.(2018)]{2018arXiv180310781C} Cai, Z., Hamden, E., Matuszewski, M., et al.\ 2018, arXiv:1803.10781
\bibitem[Chapman et al.(2004)]{2004ApJ...606...85C} Chapman, S.~C., Scott, D., Windhorst, R.~A., et al.\ 2004, \apj, 606, 85
\bibitem[Cristiani \& Vio(1990)]{1990A&A...227..385C} Cristiani, S., \& Vio, R.\ 1990, \aap, 227, 385
\bibitem[Dawson et al.(2013)]{2013AJ....145...10D} Dawson, K.~S., Schlegel, D.~J., Ahn, C.~P., et al.\ 2013, \aj, 145, 10
\bibitem[Dey et al.(2005)]{2005ApJ...629..654D} Dey, A., Bian, C., Soifer, B.~T., et al.\ 2005, \apj, 629, 654
\bibitem[Dong et al.(2008)]{2008MNRAS.383..581D} Dong, X., Wang, T., Wang, J., et al.\ 2008, \mnras, 383, 581
\bibitem[Erb et al.(2006)]{2006ApJ...647..128E} Erb, D.~K., Steidel, C.~C., Shapley, A.~E., et al.\ 2006, \apj, 647, 128
\bibitem[Fardal et al.(2001)]{2001ApJ...562..605F} Fardal, M.~A., Katz, N., Gardner, J.~P., et al.\ 2001, \apj, 562, 605
\bibitem[Fathivavsari et al.(2015)]{2015MNRAS.454..876F} Fathivavsari, H., Petitjean, P., Noterdaeme, P., et al.\ 2015, \mnras, 454, 876
\bibitem[Fathivavsari et al.(2016)]{2016MNRAS.461.1816F} Fathivavsari, H., Petitjean, P., Noterdaeme, P., et al.\ 2016, \mnras, 461, 1816
\bibitem[Fathivavsari et al.(2017)]{2017MNRAS.466L..58F} Fathivavsari, H., Petitjean, P., Zou, S., et al.\ 2017, \mnras, 466, L58
\bibitem[Fathivavsari et al.(2018)]{2018MNRAS.tmp..972F} Fathivavsari, H., Petitjean, P., Jamialahmadi, N., et al.\ 2018, \mnras,
\bibitem[Finley et al.(2013)]{2013A&A...558A.111F} Finley, H., Petitjean, P., P{\^a}ris, I., et al.\ 2013, \aap, 558, A111
\bibitem[Forster et al.(2001)]{2001ApJS..134...35F} Forster, K., Green, P.~J., Aldcroft, T.~L., et al.\ 2001, \apjs, 134, 35
\bibitem[Francis et al.(1991)]{1991ApJ...373..465F} Francis, P.~J., Hewett, P.~C., Foltz, C.~B., et al.\ 1991, \apj, 373, 465
\bibitem[Fynbo et al.(2010)]{2010MNRAS.408.2128F} Fynbo, J.~P.~U., Laursen, P., Ledoux, C., et al.\ 2010, \mnras, 408, 2128
\bibitem[Garnett et al.(2017)]{2017MNRAS.472.1850G} Garnett, R., Ho, S., Bird, S., \& Schneider, J.\ 2017, \mnras, 472, 1850
\bibitem[Gronwall et al.(2007)]{2007ApJ...667...79G} Gronwall, C., Ciardullo, R., Hickey, T., et al.\ 2007, \apj, 667, 79
\bibitem[Haiman et al.(2000)]{2000ApJ...537L...5H} Haiman, Z., Spaans, M., \& Quataert, E.\ 2000, \apjl, 537, L5
\bibitem[Hennawi et al.(2009)]{2009ApJ...693L..49H} Hennawi, J.~F., Prochaska, J.~X., Kollmeier, J., \& Zheng, Z.\ 2009, \apjl, 693, L49
\bibitem[Ho(2005)]{2005ApJ...629..680H} Ho, L.~C.\ 2005, \apj, 629, 680
\bibitem[Jiang et al.(2016)]{2016ApJ...821....1J} Jiang, P., Zhou, H., Pan, X., et al.\ 2016, \apj, 821, 1
\bibitem[Kennicutt et al.(2008)]{2008ApJS..178..247K} Kennicutt, R.~C., Jr., Lee, J.~C., Funes, J.~G., et al.\ 2008, \apjs, 178, 247-279
\bibitem[Kishimoto et al.(2001)]{2001ApJ...547..667K} Kishimoto, M., Antonucci, R., Cimatti, A., et al.\ 2001, \apj, 547, 667
\bibitem[Kishimoto et al.(2008)]{2008Natur.454..492K} Kishimoto, M., Antonucci, R., Blaes, O., et al.\ 2008, \nat, 454, 492
\bibitem[Kulkarni et al.(2006)]{2006ApJ...636...30K} Kulkarni, V.~P., Woodgate, B.~E., York, D.~G., et al.\ 2006, \apj, 636, 30
\bibitem[Kulkarni et al.(2012)]{2012ApJ...749..176K} Kulkarni, V.~P., Meiring, J., Som, D., et al.\ 2012, \apj, 749, 176
\bibitem[Leclercq et al.(2017)]{2017A&A...608A...8L} Leclercq, F., Bacon, R., Wisotzki, L., et al.\ 2017, \aap, 608, A8
\bibitem[Leitherer et al.(1999)]{1999ApJS..123....3L} Leitherer, C., Schaerer, D., Goldader, J.~D., et al.\ 1999, \apjs, 123, 3
\bibitem[Liu et al.(2015)]{2015ApJS..217...11L} Liu, W.-J., Zhou, H., Ji, T., et al.\ 2015, \apjs, 217, 11
\bibitem[Lusso et al.(2015)]{2015MNRAS.449.4204L} Lusso, E., Worseck, G., Hennawi, J.~F., et al.\ 2015, \mnras, 449, 4204
\bibitem[Malkan \& Sargent(1982)]{1982ApJ...254...22M} Malkan, M.~A., \& Sargent, W.~L.~W.\ 1982, \apj, 254, 22
\bibitem[Morrissey et al.(2007)]{2007ApJS..173..682M} Morrissey, P., Conrow, T., Barlow, T.~A., et al.\ 2007, \apjs, 173, 682
\bibitem[M{\o}ller et al.(2004)]{2004A&A...422L..33M} M{\o}ller, P., Fynbo, J.~P.~U., \& Fall, S.~M.\ 2004, \aap, 422, L33
\bibitem[Nagamine et al.(2004)]{2004MNRAS.348..435N} Nagamine, K., Springel, V., \& Hernquist, L.\ 2004, \mnras, 348, 435
\bibitem[Nestor et al.(2011)]{2011ApJ...736...18N} Nestor, D.~B., Shapley, A.~E., Steidel, C.~C., \& Siana, B.\ 2011, \apj, 736, 18
\bibitem[Nestor et al.(2013)]{2013ApJ...765...47N} Nestor, D.~B., Shapley, A.~E., Kornei, K.~A., Steidel, C.~C., \& Siana, B.\ 2013, \apj, 765, 47
\bibitem[Neugebauer et al.(1987)]{1987ApJS...63..615N} Neugebauer, G., Green, R.~F., Matthews, K., et al.\ 1987, \apjs, 63, 615
\bibitem[Noterdaeme et al.(2014)]{2014A&A...566A..24N} Noterdaeme, P., Petitjean, P., P{\^a}ris, I., et al.\ 2014, \aap, 566, A24
\bibitem[Ohyama et al.(2003)]{2003ApJ...591L...9O} Ohyama, Y., Taniguchi, Y., Kawabata, K.~S., et al.\ 2003, \apjl, 591, L9
\bibitem[Pan et al.(2017)]{2017ApJ...835..218P} Pan, X., Zhou, H., Ge, J., et al.\ 2017, \apj, 835, 218
\bibitem[Prochaska et al.(2005)]{2005ApJ...635..123P} Prochaska, J.~X., Herbert-Fort, S., \& Wolfe, A.~M.\ 2005, \apj, 635, 123
\bibitem[Prochaska et al.(2014)]{2014MNRAS.438..476P} Prochaska, J.~X., Madau, P., O'Meara, J.~M., \& Fumagalli, M.\ 2014, \mnras, 438, 476
\bibitem[Scarlata et al.(2009)]{2009ApJ...704L..98S} Scarlata, C., Colbert, J., Teplitz, H.~I., et al.\ 2009, \apjl, 704, L98
\bibitem[Shapley et al.(2003)]{2003ApJ...588...65S} Shapley, A.~E., Steidel, C.~C., Pettini, M., \& Adelberger, K.~L.\ 2003, \apj, 588, 65
\bibitem[Shakura \& Sunyaev(1973)]{1973A&A....24..337S} Shakura, N.~I., \& Sunyaev, R.~A.\ 1973, \aap, 24, 337
\bibitem[Shen et al.(2011)]{2011ApJS..194...45S} Shen, Y., Richards, G.~T., Strauss, M.~A., et al.\ 2011, \apjs, 194, 45
\bibitem[Shields(1978)]{1978Natur.272..706S} Shields, G.~A.\ 1978, \nat, 272, 706
\bibitem[Nestor et al.(2013)]{2013ApJ...765...47N} Nestor, D.~B., Shapley, A.~E., Kornei, K.~A., Steidel, C.~C., \& Siana, B.\ 2013, \apj, 765, 47
\bibitem[Siana et al.(2013)]{2013AAS...22122805S} Siana, B.~D., Alavi, A., Richard, J., et al.\ 2013, American Astronomical Society Meeting Abstracts \#221, 221, 228.05
\bibitem[Siana et al.(2015)]{2015ApJ...804...17S} Siana, B., Shapley, A.~E., Kulas, K.~R., et al.\ 2015, \apj, 804, 17
\bibitem[Syphers et al.(2009)]{2009ApJS..185...20S} Syphers, D., Anderson, S.~F., Zheng, W., et al.\ 2009, \apjs, 185, 20
\bibitem[Trump et al.(2006)]{2006ApJS..165....1T} Trump, J.~R., Hall, P.~B., Reichard, T.~A., et al.\ 2006, \apjs, 165, 1
\bibitem[Vanden Berk et al.(2001)]{2001AJ....122..549V} Vanden Berk, D.~E., Richards, G.~T., Bauer, A., et al.\ 2001, \aj, 122, 549
\bibitem[Vanzella et al.(2010)]{2010MNRAS.404.1672V} Vanzella, E., Siana, B., Cristiani, S., \& Nonino, M.\ 2010, \mnras, 404, 1672
\bibitem[Vanzella et al.(2012)]{2012ApJ...751...70V} Vanzella, E., Guo, Y., Giavalisco, M., et al.\ 2012, \apj, 751, 70
\bibitem[Weidinger et al.(2004)]{2004Natur.430..999W} Weidinger, M., M{\o}ller, P., \& Fynbo, J.~P.~U.\ 2004, \nat, 430, 999
\bibitem[Wisotzki et al.(2016)]{2016A&A...587A..98W} Wisotzki, L., Bacon, R., Blaizot, J., et al.\ 2016, \aap, 587, A98
\bibitem[Wolfe et al.(1986)]{1986ApJS...61..249W} Wolfe, A.~M., Turnshek, D.~A., Smith, H.~E., \& Cohen, R.~D.\ 1986, \apjs, 61, 249
\bibitem[Wolfe et al.(2005)]{2005ARA&A..43..861W} Wolfe, A.~M., Gawiser, E., \& Prochaska, J.~X.\ 2005, \araa, 43, 861
\bibitem[York et al.(2000)]{2000AJ....120.1579Y} York, D.~G., Adelman, J., Anderson, J.~E., Jr., et al.\ 2000, \aj, 120, 1579
\bibitem[Zhang et al.(2010)]{2010ApJ...714..367Z} Zhang, S., Wang, T.-G., Wang, H., et al.\ 2010, \apj, 714, 367
\bibitem[Zhang et al.(2014)]{2014ApJ...786...42Z} Zhang, S., Wang, H., Wang, T., et al.\ 2014, \apj, 786, 42
\bibitem[Zhang et al.(2015)]{2015ApJ...815..113Z} Zhang, S., Zhou, H., Shi, X., et al.\ 2015, \apj, 815, 113
\bibitem[Zheng et al.(1997)]{1997ApJ...475..469Z} Zheng, W., Kriss, G.~A., Telfer, R.~C., Grimes, J.~P., \& Davidsen, A.~F.\ 1997, \apj, 475, 469
\end{thebibliography}
\end{document}